\documentstyle[12pt]{article}

\begin{document}
\begin{center}
{\Large \bf Generation of density perturbations
due to the birth of baryons }
\bigskip

{\large D.L.~Khokhlov}
\smallskip

{\it Sumy State University, R.-Korsakov St. 2, \\
Sumy 244007, Ukraine\\
E-mail: khokhlov@cafe.sumy.ua}
\end{center}

\begin{abstract}
Generation of adiabatic density perturbations
from fluctuations caused by the birth of baryons is considered.
This is based on the scenario of baryogenesis in which
the birth of protons takes place
at the temperature equal to the mass of electron.
\end{abstract}

The observed large scale structure of the universe forms
from the primary density perturbations after recombination
$z_{rec}=1400$~\cite{Zeld}.
In the epoch of recombination, the photons decouple
from the baryons and last scatter.
After recombination the photons do not interact with the baryons,
so the cosmic microwave background (CMB) anisotropy allows us to
define the density perturbations of the baryonic matter
in the epoch of recombination.
Fluctuations in the CMB on the large scale
detected by COBE satellite~\cite{Ben} are
$\Delta T/T=1.06\times 10^{-5}$.
Generation of adiabatic density perturbations
from quantum fluctuations is considered
within the framework of inflation cosmology~\cite{Lin}.

Let us consider generation of adiabatic density perturbations
from fluctuations caused by the birth of baryons.
According to the scenario of baryogenesis proposed in~\cite{Kh1},
at $T>m_e$,
primordial plasma consists of neutral fermions.
Neutral electrons
are in the state being the superposition of electron and
positron
\begin{equation}
|e^0>=\frac{1}{\sqrt 2}(e^-+e^+).
\label{eq:en}
\end{equation}
Neutral protons
are in the state being the superposition of proton and
antiproton
\begin{equation}
|p^0>=\frac{1}{\sqrt 2}(\bar p + p).
\label{eq:pn}
\end{equation}
At $T>m_e$,
there exists neutral proton-electron symmetry.
Proton-electron equilibrium
is defined by the proton-electron mass difference.
At $T=m_{e}$, pairs of neutral electrons annihilate into photons,
pairs of neutral protons and electrons survive as
protons and electrons.
At $T=m_{e}$, the baryon-photon ratio is given by
\begin{equation}
\frac{N_{b}}{N_{\gamma}}=\frac{3}{4}
\left(\frac{1}{2}\right)^{5}\left(\frac{m_{e}}{m_{p}}\right)^{2}.
\label{eq:NbNg}
\end{equation}
Calculations yield $N_{b}/N_{\gamma}=6.96\times 10^{-9}$.
It should be noted that the observed value of $N_{b}/N_{\gamma}$ lies
in the range $2-15\times 10^{-10}$~\cite{ISSI}.
Possible explanation of the discrepancy between the result of
calculations and the observed value is that the most fraction of
baryonic matter
decays into non-baryonic matter during the evolution of the universe.

The birth of baryons causes potential fluctuations
\begin{equation}
\frac{\delta\varphi}{\varphi}=\frac{\rho_{b}}{\rho}.
\label{eq:del}
\end{equation}
In the case of homogeneous universe, the spectrum of fluctuations
is flat.
At $T=m_{e}$,
the value of fluctuations (\ref{eq:del}) is given by
\begin{equation}
\frac{\delta\varphi}{\varphi}=\frac{4}{3}\times\frac{7}{8}
\frac{N_{b}}{N_{\gamma}}\frac{m_{p}}{m_{e}}.
\label{eq:del1}
\end{equation}
Here the factor $4/3\times 7/8$ describes transition from
the particle number density to the energy density.
Calculations yield $\delta\varphi/\varphi=1.49 \times 10^{-5}$.

Density perturbations are related to potential fluctuations
via the Poisson equation.
Let us assume that the state of the fluid in the universe is given by
\begin{equation}
|\psi>=\frac{1}{\sqrt{3}}(|\psi_x>+|\psi_y>+|\psi_z>).
\label{eq:psi}
\end{equation}
In comparison with the separate particles
$|\psi_x>$, $|\psi_y>$, $|\psi_z>$,
for the particles being in the superpositional state (\ref{eq:psi}),
the probability density is three times greater.
In this case the adiabatic speed of sound is given by
\begin{equation}
v_s=\left(\frac{3\partial p}{\partial\rho}\right)^{1/2}.
\label{eq:vs}
\end{equation}
For radiation with the equation of state
\begin{equation}
p=\frac{\rho c^2}{3}
\label{eq:prho}
\end{equation}
the adiabatic speed of sound is equal to the speed of light
\begin{equation}
v_s=c.
\label{eq:vs1}
\end{equation}
At $T=m_{e}$,
the fluid is radiation dominated.
Relation between potential fluctuations and density perturbations
of radiation is given by
\begin{equation}
\frac{\delta\varphi}{\varphi}=\frac{\delta\rho_{\gamma}}{\rho_{\gamma}}.
\label{eq:dphi}
\end{equation}
Relation between density perturbations of the baryonic matter
and density perturbations of radiation is given by
\begin{equation}
\frac{\delta\rho_{b}}{\rho_{b}}=\frac{3}{4}
\frac{\delta\rho_{\gamma}}{\rho_{\gamma}}.
\label{eq:drn}
\end{equation}
In view of (\ref{eq:vs}), fluctuations in the CMB
is given by
\begin{equation}
\frac{\delta T}{T}=\frac{\delta\rho_{b}}{\rho_{b}}.
\label{eq:dtt}
\end{equation}
Calculations yield $\delta T/T=1.12 \times 10^{-5}$.

\end{document}